\documentclass[twocolumn,showpacs,preprintnumbers,amsmath,amssymb]{revtex4}

\usepackage{graphicx}
\usepackage{dcolumn}
\usepackage{bm}

\begin{document}


\title{Microscopic Structural Relaxation in a Sheared Supercooled
Colloidal Liquid}

\author{Dandan Chen$^1$}
\email{dchen@physics.emory.edu}
\author{Denis Semwogerere$^1$}
\author{Jun Sato$^2$}
\author{Victor Breedveld$^2$}
\author{Eric R.~Weeks$^1$}%
\affiliation{$^1$ Department of Physics, Emory University, Atlanta, GA 30322 }
\affiliation{$^2$ School of Chemical and Biomolecular Engineering, Georgia Institute of Technology, 311 Ferst Drive NW, Atlanta, Georgia 30332-0100 }

\date{\today}

\begin{abstract}
The rheology of dense amorphous materials under large shear strain
is not fully understood, partly due to the difficulty of directly
viewing the microscopic details of such materials.  We use a
colloidal suspension to simulate amorphous materials, and study the
shear-induced structural relaxation with fast confocal microscopy.
We quantify the plastic rearrangements of the particles in several
ways.  Each of these measures of plasticity reveals spatially
heterogeneous dynamics, with localized regions where many particles
are strongly rearranging by these measures.  We examine the shapes
of these regions and find them to be essentially isotropic,
with no alignment in any particular direction.  Furthermore,
individual particles are equally likely to move in any direction,
other than the overall bias imposed by the strain.
\end{abstract}


\maketitle

\section{Introduction}
\label{Introduction}

Many common materials have an amorphous structure,
such as shaving cream, ketchup, toothpaste, gels, and window glass
\cite{Angell00,Coussot05,Ubbink08,Mezzenga05}.  In some situations
these are viscous liquids, for example when window glass is heated
above the glass transition temperature, or a shaving cream foam that
has been diluted by water to become a liquid with bubbles in it.
In other situations these are viscoelastic or elastic solids,
such as gels and solid window glass \cite{Liu98}.  For solid-like
behavior, when a small stress is applied, the materials maintain
their own shapes; at larger stresses above the yield stress,
they will start to flow \cite{Coussot02,Bonn05,Merkt04}.
Understanding how these materials yield and flow is important for
the processing of these materials, and understanding their
strength in the solid state \cite{Spaepen77,Ma07,Schuh07}.

A particularly interesting system to study is a colloidal
suspension.  These consist of micron or submicron
sized solid particles in a liquid.  At high particle
concentration, macroscopically these are pastes and thus
of practical relevance \cite{Coussot02}.  Additionally, for
particles with simple hard-sphere like interactions, colloidal
suspensions also serve as useful model systems of liquids,
crystals, and glasses \cite{Pusey86,Haw02,Suresh06,Schall07}.
Such colloidal model systems have the advantage that they can
be directly observed with microscopy
\cite{habdas02,prasad07,Elliot01}.  Our particular interest
in this paper is using colloidal suspensions to model supercooled
and glassy materials.  The control parameter for hard sphere
systems is the
concentration, expressed as the volume fraction $\phi$, and
the system acts like a glass for $\phi > \phi_g \approx 0.58$.
The transition is the point where particles no longer diffuse
through the sample; for $\phi < \phi_g$ spheres do diffuse
at long times, although the asymptotic diffusion coefficient
$D_\infty$ decreases sharply as the concentration increases
\cite{vanmegen91,speedy98,brambilla09}.  The transition
at $\phi_g$ occurs even though the spheres are not completely
packed together; in fact, the density must be increased to
$\phi_{\rm RCP} \approx 0.64$ for ``random-close-packed'' spheres
\cite{ohern03,bernal64,torquato00,torquato04,ohern04} before the
spheres are motionless.
Prior work has shown remarkable similarities
between colloidal suspensions and conventional molecular glasses
\cite{Pusey86,vanmegen91,vanblaaderen95,vanmegen93,vanmegen94,snook91,
bartsch93,bartsch95,mason95glass}.

One important unsolved problem related to amorphous materials is to
understand the origin of their unique rheological behavior under
shear flow.  Early in the 70's, theory predicted the existence of
``flow defects'' beyond yielding \cite{Spaepen77}, later termed
shear transformation zones (STZ) \cite{Falk98}.  These
microscopic motions result in plastic deformation of the sheared
samples \cite{goyon08,bocquet09}.
Simulations later
found STZs by examining the microscopic local particle motions
\cite{Schuh07,Yamamoto97,Yamamoto04}.  Recently fast confocal
microscopy has been used to examine the shear of colloidal
suspensions \cite{Besseling07}, and STZ's have been directly
observed \cite{Schall07}.  This provided direct evidence to support
theoretical work on STZ's \cite{Spaepen77,Maeda81,Falk98,Shi05}.

However, questions still remain.  First, most of the prior work
has focused on the densest possible samples, at concentrations
which are glassy ($\phi > \phi_G$) \cite{Besseling07,Schall07}.
Given that the macroscopic viscosity of colloidal suspensions
change dramatically near and above $\phi_G$ \cite{cheng02},
it is of interest to study slightly less dense suspensions
under shear, for which rearrangements might be easier
\cite{Weeks00}.  In this paper, we present such results.  Second,
prior investigations of sheared amorphous materials have used
a variety of different ways to quantify plastic deformation
\cite{Wang06,Utter08,Besseling07,Schall07}.  In this paper,
we will compare and contrast plastic deformations defined in
several different ways.  While they do capture different aspects
of plastic deformation, we find that some results are universal.
In particular, in a sheared suspension, there are three distinct
directions:  the strain velocity, the velocity gradient,
and the direction mutually perpendicular to the first two (the
``vorticity'' direction).  We find that plastic deformations are
isotropic with respect to these three directions, apart from the
trivial anisotropy due to the velocity gradient.  The deformations
are both isotropic in the sense of individual particle motions,
and in the sense of the shape of regions of rearranging particles.

\section{Experimental Methods}
\label{method}

\begin{figure}[htbp]
\includegraphics[scale=0.35,angle=-90]{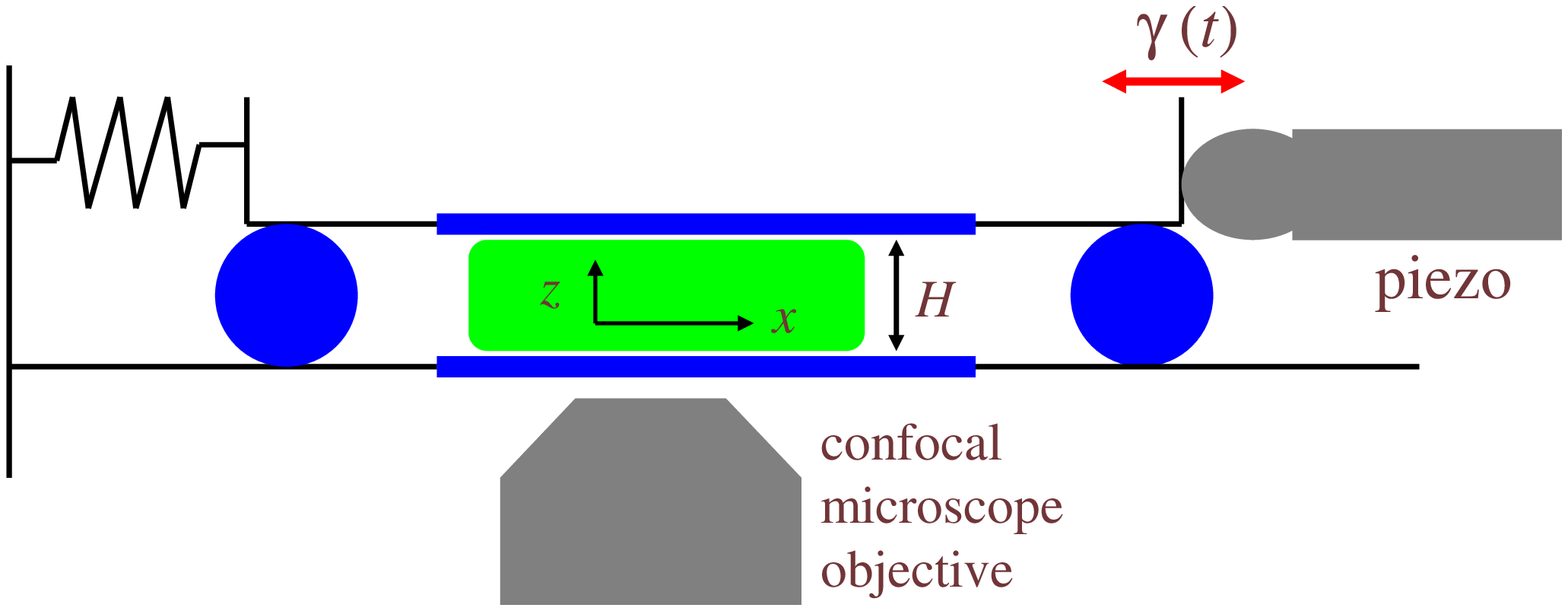}
\caption{\label{ShearCell}
(Color online)
Sketch of the shear cell.
A fluorescent sample (gray, green online)
is put between two parallel glass plates 
(dark gray, blue online)
with gap $H$ set by three ball bearings, two of which are shown.
The top plate is movable,
controlled by a piezo motor driven by a triangular wave.
The bottom plate is fixed,
and the confocal microscope takes images from underneath.
Note the definition of the coordinate system, where $x$ is in the
velocity direction and $z$ is in the velocity gradient direction.
}
\end{figure}

The experimental setup of our shear cell is shown in
Fig.~\ref{ShearCell} and is similar to that described in
Refs.~\cite{Petekidis02,Besseling07}.
The glass plates are 15~mm in diameter, and to the top plate is glued
a small piece of glass with dimensions 5~mm$\times$1~mm,
with the long dimension oriented in the direction of motion $x$.
The purpose of this piece of glass is to decrease the effective gap size.
Between the plates are three ball bearings,
used to control the gap size; for all of our data, we maintain a
gap size of $H=130$~$\mu$m.
Over the 5~mm length of the small pieces
of glass, the gap varies by no more than 15~$\mu$m; over the
narrower dimension of the small pieces of glass, the gap varies
by no more than 10~$\mu$m.  Thus, overall the sample is between
two plates which are parallel to within 1\%
and in the direction of shear, they are parallel to within 0.3\%.

A droplet of the sample (volume $\sim$200~$\mu$l) is placed between the
two pieces of glass.  The top plate is free to move in the $x$
direction, and the bottom plate is motionless.  The shear rate
is controlled by a piezo-electric actuator (Piezomechanik GmbH
Co.)  driven by a triangular wave signal with a period ranging
from $T=150$ to 450~s and an amplitude of $A=175$~$\mu$m.  Thus
we achieve strains of $\gamma_0 = A/H = 1.4$.  Prior to taking
data, we allow the shear cell to go through at least one complete
period, but usually not more than three complete periods.

\begin{figure}[htbp]
\includegraphics[scale=0.35]{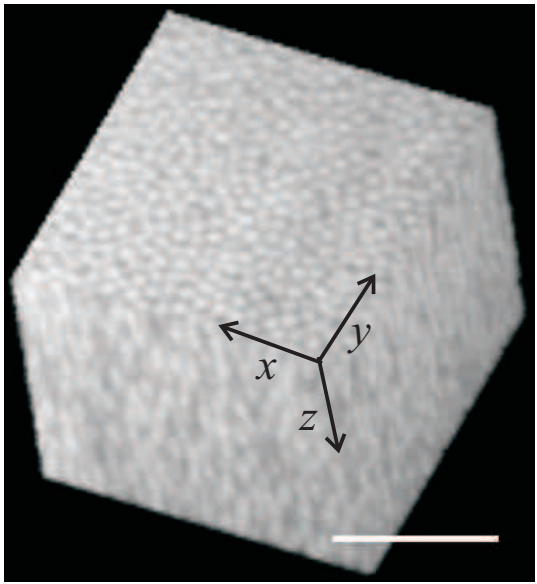}
\caption{\label{Colloids}
A 20$\times$20$\times$20~$\mu$m$^3$ image of a supercooled colloidal liquid
taken 15~$\mu$m away from the fixed bottom plate
by our confocal microscope in less than 1~s.
The scale bar represents 10~$\mu$m.
}
\end{figure}

Our samples (Fig.~\ref{Colloids}) are poly(methyl methacrylate)
colloids sterically stabilized with poly-12-hydroxystearic acid
\cite{Antl86}.  These particles are suspended in a mixture of
85\% cyclohexybromide and 15\% decalin by weight.
This mixture matches both the density and the index of
refraction of the particles.  To visualize the particles,
they are dyed with rhodamine $6G$ \cite{Dinsmore01}.  The
particles have a radius $a = 1.05 \pm 0.04$~$\mu$m, with the
error bar indicating the uncertainty in the mean diameter;
additionally the particles have a polydispersity of no more than 5\%
(as these particles can crystallize fairly easily)
\cite{Pusey86,Auer01,Henderson96,Megen07}.  

In this work we study several samples with volume fractions $\phi$
between 0.51 and 0.57.  Thus, our samples are quite dense liquids,
comparable to prior work with ``supercooled'' colloidal liquids
\cite{Pusey86,Weeks00}.  The differences in volume fraction between
samples are certain to within $\pm 0.01$, and the absolute volume
fraction has a systematic uncertainty of $\pm 0.06$ due to the
uncertainty of the particle radius $a$.  However, none of our
samples appear glassy, and thus we are confident our maximum
volume fraction is less than $\sim 0.6$.  While the particles in
decalin behave as hard-spheres, in our solvent mixture they carry
a slight charge.  This does not seem to affect the phase behavior
dramatically at high volume fractions such as what we consider in
this work; see for example \cite{Gasser01,Hernandez-Guzman09}.


To characterize the relative importance of Brownian motion and the
imposed strain field, we can compute the modified Peclet number,
$Pe^{\star} = \dot{\gamma} a^2/2D_{\infty}$, where $D_{\infty}$
is the long time diffusion coefficient of the quiescent sample.
We measure $D_{\infty}$ from mean square displacement data taken
from the quiescent sample with the same volume fraction.  The large
$\Delta t$ data for the mean square displacement can be fit using
$\langle \Delta x^2 \rangle = 2 D_\infty \Delta t$.  Roughly,
$a^2/2 D_{\infty}$ reflects the average duration a particle is
caged by its neighbors in the dense suspension.

$\dot{\gamma}$ is the imposed strain rate
$2A/(HT)$, and $a$ is the particle size. 
The extra factor
of 2 in $\dot{\gamma}$ is because we use a triangle wave,
and thus the half period sets the strain rate.  For our
samples, we find $D_{\infty} \approx 5 \times 10^{-4}$~$\mu$m$^2/$s, and we
have $\dot{\gamma}$ ranging from 0.0060 to 0.0180 s$^{-1}$;
thus $Pe^{\star} \approx 7 - 20$.
Given that $Pe^{\star} > 1$,
the implication is that
the motions we will observe are primarily caused by the strain,
rather than due to Brownian motion.  We use the modified Peclet
number based on $D_{\infty}$ rather than the bare Peclet number
based on the dilute-limit diffusivity $D_0$, as we will focus
our attention on the dynamics at long time scales, which we will
show are indeed shear-induced.


Shear-induced crystallization has been found in previous
work \cite{Haw98,Duff07}.  As we wish to focus on the case
of sheared amorphous materials, we check our data to look for
crystalline regions using standard methods which detect ordering
\cite{Gasser01,Hernandez-Guzman09,Dullens06,Frenkel96,Steinhardt83}.
Using these methods, we find that particles in apparently
crystalline regions comprise less than 3\% of the particles in
each of our experiments, and are not clustered, suggesting that
the apparently crystalline regions are tiny.  This confirms that
our samples maintain amorphous structure over the time scale of
our experiments, although perhaps if we continued the shearing
over many more cycles we would find shear-induced crystallization.

We use a confocal microscope to image our sample (the ``VT-Eye'',
Visitech), using a 100$\times$ oil lens (numerical aperture =
1.40) \cite{vanblaaderen92,Dinsmore01,prasad07}.  A 3D image with
a volume 50$\times$50$\times$20~$\mu$m$^3$ is acquired in less
than 2~s; these images contain about 6000 particles.  The 3D
image is $256 \times 256 \times 100$ pixels, so approximately
0.2~$\mu$m per pixel in each direction.  Figure \ref{Colloids}
shows a representative image from a somewhat smaller volume.
The 2~s acquisition time is several orders of magnitude faster than
the diffusion for particles in our high volume fraction sample.
To avoid any boundary effects \cite{Shereda08}, we scan a volume at
least 20~$\mu$m away from the bottom plate.  Particle positions
are determined with an accuracy of $0.05$~$\mu$m in $x$ and
$y$, and $0.1$~$\mu$m in $z$.  This is done by first spatially
filtering the 3D image with a bandpass filter designed to remove
noise at high and low spatial frequencies, and then looking for
local maxima in the image intensity \cite{Crocker96}.
Our tracking algorithm is similar to prior work
\cite{Crocker96,Dinsmore01,Besseling09}, where we first identify
particles within each 3D image, next remove the overall average
shear-induced motion from all of the particles, then track
the particles in the ``co-shearing'' reference frame using
conventional techniques \cite{Crocker96}, and finally add
back in the shear-induced motion that was previously removed.
This is similar to the ``Iterated CG tracking'' method described
in Ref.~\cite{Besseling09}.  The key idea of this tracking is
that particles should not move more than an inter-particle
spacing between each image; this condition is satisfied in the
co-shearing reference frame.

Due to the strain, particles that start near one face of the
imaging volume are carried outside the field of view, while on
the opposite face new particles are brought inside.  Thus, for
larger strains, the total number of particles viewed for the
entire duration diminishes.  For the data discussed in this work,
we consider both instantaneous quantities and quantities averaged
over the entire half-cycle of strain.  For the former, we view
$\sim 5500$ particles, while for the latter, we typically can
follow $\sim 3000$ particles, which limits our statistics
slightly.

\section{Results}
\label{Results}

\subsection{Locally observed strain}

Our goal is to understand if the local shear-induced motion is
isotropic in character.  However, first we seek to understand and
quantify the more global response of our sheared samples.

\begin{figure}[htbp]
\includegraphics[scale=0.35, angle=-90]{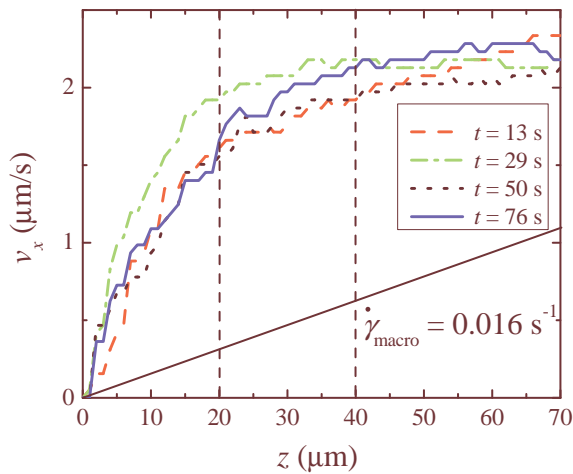}
\caption{\label{ShearBand}
(Color online)
Shear profile $v_x(z)$ within
a 50$\times$50$\times$70~$\mu$m$^3$ volume,
measured at different times during
one half period $T/2=84$~s.  The total macroscopic strain applied in
84~s is 1.35.  The sample has a volume fraction $\phi \approx
0.50$, and the global applied shear rate is $\dot{\gamma}_{\rm
macro}=0.016$~s$^{-1}$ controlled by the piezo motor.  If there
was not a shear band, one would expect to observe a linear
velocity profile as indicated by the lower thick diagonal line.
The four curves represent different times during the half
period as indicated.  Note that these data are obtained using a
coarse-grained image velocimetry method \cite{Besseling09},
rather than particle tracking.  Here we take rapid images over
the full volume with a spacing of $z=1$~$\mu$m in the vertical
direction.  For each value of $z$, we cross-correlate subsequent
images to obtain a mean instantaneous velocity $v_x$ with a
resolution of 0.05~$\mu$m/s, set by the pixel size and the time
between images.  See
Ref.~\cite{Besseling09} for further details.
}
\end{figure}

When shearing disordered materials or complex fluids,
one often finds shear localization or shear banding, due
to the nonlinear yielding and relaxation in local regions
\cite{Bonn05,Coussot02,Chan04,Lauridsen04}.
To check for this in our data, we start by taking 3D images
under the applied shear rate $\dot{\gamma}_{\rm macro}=0.016$~s$^{-1}$
over a very large range in $z$, from 0 to 70~$\mu$m
away from the bottom plate, almost half of the gap between two
shearing plates.  To allow us to visualize more clearly over
such a large depth, we dilute the dye concentration by mixing
dyed and undyed colloids at a number ratio of around 1$:$80
and keeping the desired volume fraction $\phi \approx 0.50$.
Our sample does indeed form a shear
band, as shown in Fig.~\ref{ShearBand}, which shows the
particle velocity $v_x$ in the direction of the shear as a
function of the depth $z$.  The velocity changes rapidly with
$z$ in the range $0 < z < 20$~$\mu$m, and then more slowly
for $z > 20$~$\mu$m; thus much of the shear occurs adjacent
to the stationary plate at $z=0$~$\mu$m, similar to prior work
which found shear adjacent to one of the walls
\cite{Cohen06,Bonn05,Coussot02,Chan04,Lauridsen04}.
Furthermore, the
velocity profile is relatively stable during the course of the
half-period, as seen by the agreement between the velocity
profiles taken at different times during this half period
(different lines in Fig.~\ref{ShearBand}.  Thus, the shear band
develops quickly inside the supercooled colloidal liquid, and
remains fairly steady under the constant applied strain rate.
The location and size of the shear band varies from experiment to
experiment.

%

Given the existence of a shear band, the applied strain is not
always the local strain.  In this work we wish to focus on the
motion induced by a local strain, rather than the global formation
of shear bands.  Thus, for all data sets presented below, we always
calculate the local instantaneous strain rate $\dot{\gamma}_{\rm
meso}(t) = \frac{v_x (z+ \Delta z,t) - v_x (z,t)}{\Delta z}$.
Here $\Delta z = 20$~$\mu$m is the height
of the imaged volume.  Related to $\dot{\gamma}_{\rm meso}$,
we can calculate the total local applied strain by integrating
$\dot{\gamma}_{\rm meso}(t)$:
  \begin{equation}
  \label{totstrain}
    \gamma_{\rm meso}(t) = \int^{t}_0 \dot{\gamma}_{\rm
    meso}(t')dt'.
  \end{equation}
Furthermore, we verify that for each data set
considered below, $v_x(z)$ varies linearly with $z$ within the
experimental uncertainty, and thus $\dot{\gamma}_{\rm meso}$ is
well-defined, even if globally it varies (Fig.~\ref{ShearBand}).

\begin{figure}[htbp]
\includegraphics[scale=0.35, angle=-90]{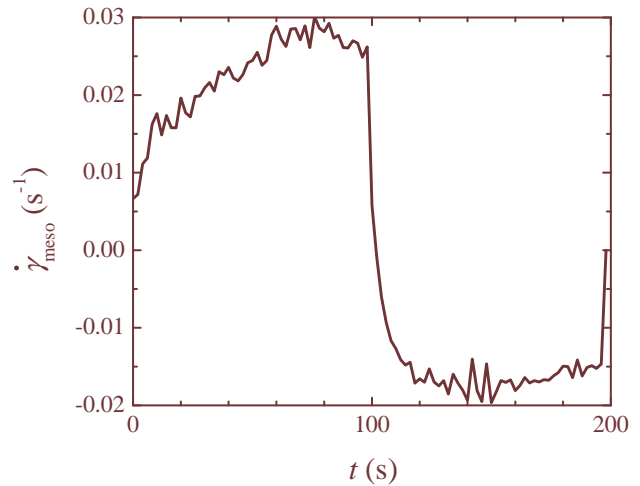}
\caption{\label{Gammadot}
A typical example of a measured local instantaneous shear rate
over one period of shearing
in a gap of 20$-$40~$\mu$m away from the fixed bottom plate.  The
sample has volume fraction
$\phi = 0.51$, the applied strain rate is
$\dot{\gamma}_{\rm macro}=0.013$~s$^{-1}$, and the period is
$T=200$~s.
}
\end{figure}

While the shape of the shear band is essentially constant
(Fig.~\ref{ShearBand}), in many cases the local strain rate
varies slightly with time.  As our forcing is a triangle wave,
over any given half cycle the global applied strain rate is
a constant.  We can measure the local strain rate for each
data set; a typical example is shown in Fig.~\ref{Gammadot}.
This figure shows the local instantaneous measured strain
rate within the region $20$~$\mu$m~$<z<40$~$\mu$m, over one full
period.  After the shearing starts, $\dot{\gamma}_{\rm meso}$
quickly rises up to 0.015~s$^{-1}$, implying that the shear
band has formed.  Subsequently, the local strain rate
continues to increase up to 0.030~s$^{-1}$ over the rest of
the half period.  The small fluctuations of $\dot{\gamma}_{\rm
meso}$ are due to the microscopic rearrangements of particles,
which can be somewhat intermittent. Given that the local
instantaneous strain rate is not constant (despite the constant
applied strain rate), we will characterize our data sets by the
time averaged local strain rate
defined as
  \begin{equation}
    \bar{\dot{\gamma}}_{\rm meso}= \frac{\gamma_{\rm meso}(t)}{t}
  \end{equation}
typically using $t=T/2$, the half period of the strain.  That is, we
consider $\bar{\dot{\gamma}}_{\rm meso}$ as one key parameter
characterizing each data set, although we will show that we see
little dependence on this parameter.  In the rest of the paper,
the choice $t=T/2$ will be assumed, except where noted when we
wish to characterize the mesoscopic strain for time scales shorter than
$T/2$.

At this point, we have defined the key control parameters, which
are measured from each experimental data set:  the strain rate
$\bar{\dot{\gamma}}_{\rm meso}$ and the strain amplitude 
$\gamma_{\rm meso}$.  We next consider how these variables
relate to the magnitude of the shear-induced particle motion.

\subsection{Individual particle motions}

\begin{figure}[htbp]
\includegraphics[scale=0.40]{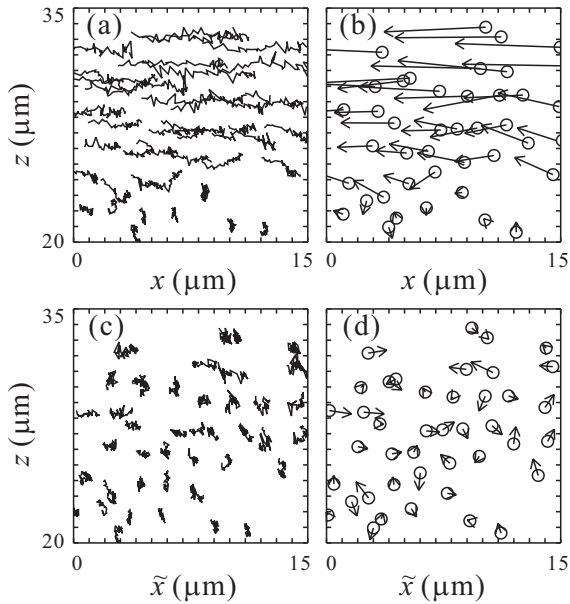}
\caption{\label{Trajectory}
Trajectories of colloids in an $xz$ slice (5~$\mu$m thick in the
$y$ direction), see Fig.~\ref{ShearCell} for the
coordinate axes.  The sample has
$\phi=0.51$, and the data shown correspond to a locally measured
accumulated strain
$\gamma_{\rm meso}=0.43$ over 45~s of data, so the
effective strain rate is $\bar{\dot{\gamma}}_{\rm
meso} = \gamma_{\rm meso}/\Delta t = 0.0096$~s$^{-1}$.
(a) Trajectories in a reference frame co-moving with the average
velocity $v_x(z=20$~$\mu$m$)$.  (b) Displacements corresponding
to data from panel (a), where the start point of each particle is
marked with a circle and the end point is marked with an arrowhead.
(c) The same data, but with the affine motion removed;
this is the $\tilde{x}z$ plane.  (d) Displacements corresponding
to panel (c).
}
\end{figure}

Because of our large local strains (measured to be 
$\gamma_{\rm meso} > 0.3$ for all cases), we observe significant
particle motion, as shown Fig.~\ref{Trajectory}(a,b).
In the laboratory reference frame, the microscopic
velocity gradient is obvious either in the raw trajectories
[Fig.~\ref{Trajectory}(a)] or in the large displacements
[Fig.~\ref{Trajectory}(b)] measured between the beginning
and end of the half period.  However, in a sense, much of
this motion is ``trivial''; we wish to observe what nontrivial local
rearrangements are caused by the strain.  To do this, we
consider the non-affine motion by removing averaged particle
displacements at the same depth $z$ from the real trajectories
of particles
\cite{Besseling07,Wang06,Lemaitre07,Yamamoto04,Maloney08},
   \begin{equation}
      \label{nonaffineeqn}
      \tilde{x}_i (t) = x_i (t) - \int^t_0 \dot{\gamma}_{\rm meso} (t')z_i (t')dt'
   \end{equation}
where the removed integral represents the shearing history of
the particle $i$.  To be clear, the shearing history is based
on the average motion within the entire imaged region, and the
remaining motion of particle $i$ is caused by interactions with
neighboring particles due to the shear.  This motion is shown in
Fig.~\ref{Trajectory}(c), showing the $\tilde{x}z$ plane rather than
the $xz$ plane; the particles move shorter distances.  Their overall
displacements in this ``de-sheared'' reference frame are shown in
Fig.~\ref{Trajectory}(d).  A few trajectories are up to 2~$\mu$m
long, comparable to the particle diameter.  These non-affine
displacements shown in Fig.~\ref{Trajectory}(d) are much smaller
than the raw displacements of Fig.~\ref{Trajectory}(b), but much
larger than thermally activated Brownian motion, which takes
more than 1000~s to diffuse over a 1~$\mu$m distance in our dense
samples (comparable to the particle radius $a$).  These non-affine
motions $\tilde{x}$ reflect shear-induced plastic changes inside
the structure.

\begin{figure}[htbp]
\includegraphics[scale=0.70, angle=-90]{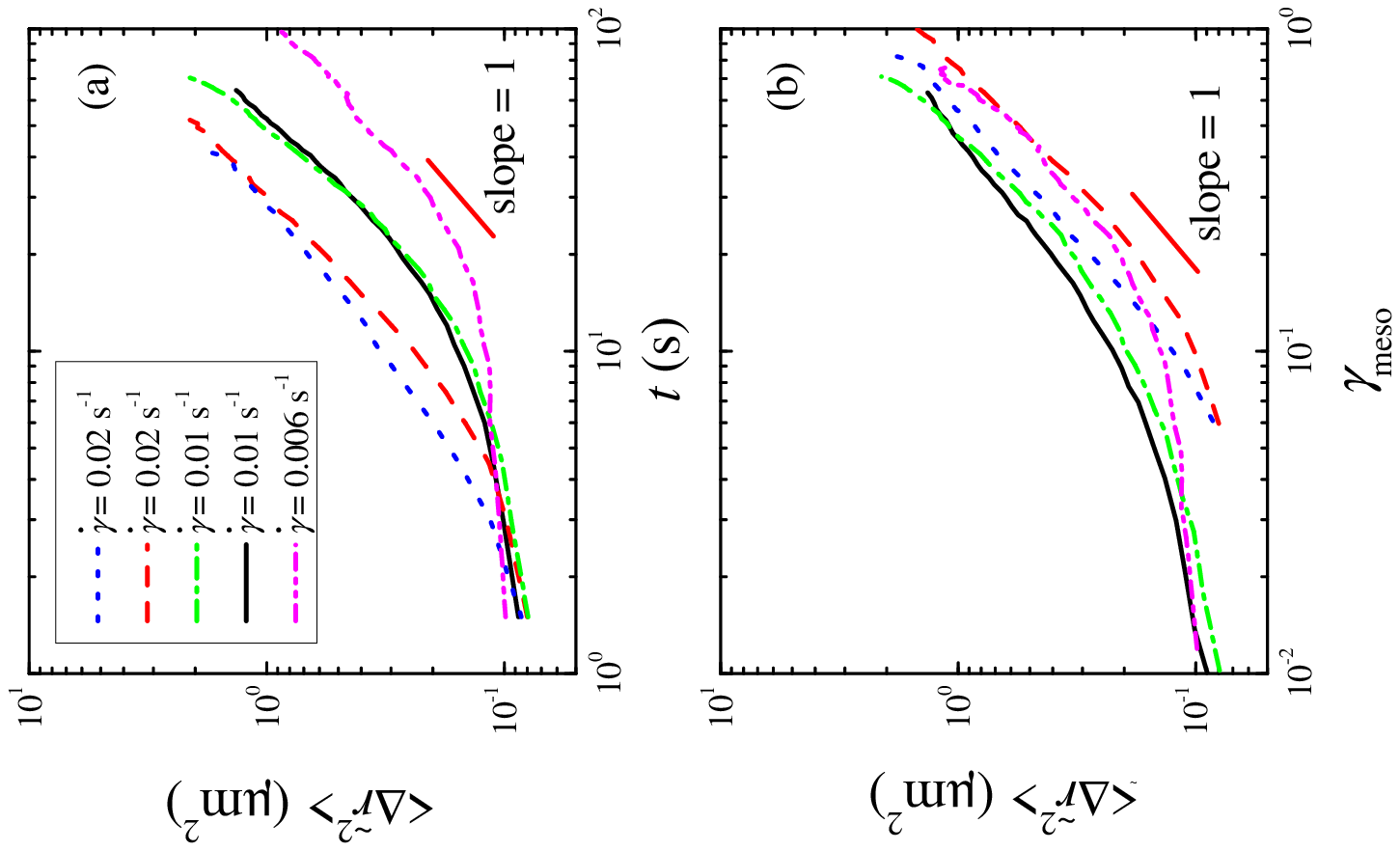}
\caption{\label{MSD}
(Color online)
(a) The mean square displacement of the non-affine motion $\Delta
\tilde{r}$ as a function of the time $t$ since the start of shear.
(b) The same data plotted as a function of the accumulated strain
$\gamma_{\rm meso} = \int^t_0 \dot{\gamma}_{\rm meso} (t')dt'$.
The five curves represents five data sets
with the same volume fraction $\phi = 0.51$
but three different shear rates $\bar{\dot{\gamma}}_{\rm meso}$ as
indicated in the figure.  The total accumulated strain is the
final point reached by each curve in panel (b).
}
\end{figure}

To quantify the amount of this non-affine motion $\tilde{x}$,
one could calculate the mean squared displacements (MSD)
often defined as
 \begin{equation}
   \langle \Delta \tilde{x}^2 \rangle  (\Delta t) = \langle
   (\tilde{x}_i(t+\Delta t)-\tilde{x}_i (t))^2 \rangle _{i,t}
 \end{equation}
where the angle brackets indicate an average over time
$t$, as well as particles $i$.  Thus this identity assumes
that environmental conditions remain the same for all the
time, since it does not depend on $t$.  However, as shown
by Fig.~\ref{Gammadot}, the shear rate $\dot{\gamma}_{\rm
local}(t)$ depends on the time.  Therefore, we use an alternate
formulation
 \begin{equation}
  \langle \Delta \tilde{x}^2\rangle (t) = \langle
  (\tilde{x}_i(t)-\tilde{x}_i (t=0))^2 \rangle _i
 \end{equation}
where the angle brackets only indicate an average over
particles, and $t$ is the time since the start of a half period
of shear.  Figure~\ref{MSD}(a) shows mean squared displacement
(MSD) of the non-affine motion $\Delta \tilde{r}^2 = \Delta
\tilde{x}^2 + \Delta y^2 + \Delta z^2$ as a function of $t$
for five different experiments with different strain rates
$\bar{\dot{\gamma}}_{\rm meso}$, from 0.02 to 0.006~s$^{-1}$.
In each case the curves nearly reach a slope of 1 on the
log-log plot, indicating that shear quickly facilitates
particles' rearrangements.  The magnitude of the motion is
$\Delta \tilde{r}^2 \approx 1$~$\mu$m$^2$, indicating that the
original structure is mostly lost \cite{Weeks02}.  

Figure~\ref{MSD}(a) also shows that $\Delta \tilde{r}^2$ is
larger for faster strain rates at the same $t$.  We find that
this motion is determined by the accumulated strain, as shown in
Fig.~\ref{MSD}(b) by replotting the MSD as a function of 
$\gamma_{\rm meso}$ (Eqn.~\ref{totstrain}).  In this graph, the
curves are grouped closer together and there is no obvious dependence on
$\bar{\dot{\gamma}}_{\rm meso}$.

It suggests that the accumulated strain is an important parameter in
the structural relaxation, which was also found in previous work on
shear transformation zones \cite{Delogu08,Maloney08}, and is similar
behavior to that seen for athermal sheared systems \cite{Pine05}.
Additionally, Fig.~\ref{MSD}(b) shows that the slopes of the curves
are close to 1 when $\gamma_{\rm meso} > 0.1$, confirming
that the accumulated strain in our experiments is large enough to
rearrange the original structure in a supercooled colloidal liquid.
We stress that the rough agreement between the curves seen in
Fig.~\ref{MSD}(b) is based on the locally measured applied strain,
and not the macroscopically applied strain.

An earlier study of steady shear applied to colloidal
glasses by Besseling et al.~\cite{Besseling07} found that the
diffusion time scale $\tau$ scaled as $\dot{\gamma}^{-0.8}$, 
and simulations also found power law scaling
\cite{Yamamoto97,Yamamoto04}.
The collapse of our MSD curves [Fig.~\ref{MSD}(b)] seems
to imply $\tau \sim \dot{\gamma}^{-1.0}$.  It is possible
that the disagreement between these results is too
slight to be clear over our limited range in $\dot{\gamma}$
(less than one decade).  Also, we study supercooled fluids
whereas Ref.~\cite{Besseling07} examines colloidal glasses.
Furthermore, our maximum local strain is $\gamma_{\rm meso} =
1.6$, while Ref.~\cite{Besseling07} considers steady strain up
to a total accumulated strain of 10.  Another recent study of
sheared colloidal glasses \cite{Eisenmann09} implies a result
similar to ours, $\tau \sim \dot{\gamma}^{-1.0}$, but did not
discuss the apparent difference with Ref.~\cite{Besseling07}.


\begin{figure}[htbp]
\includegraphics[scale=0.35,angle=-90 ]{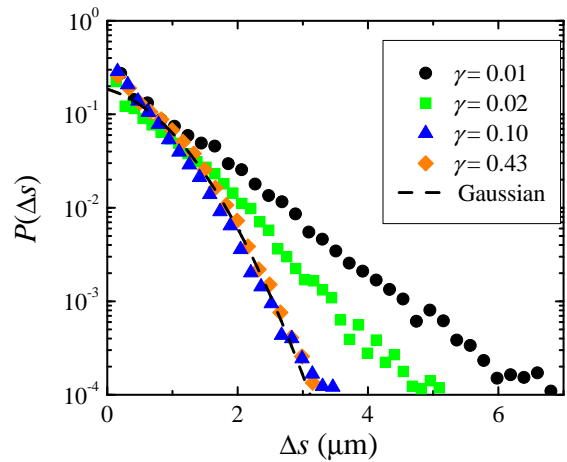}
\caption{\label{Pds}
(Color online)
Probability distribution function of
$\Delta s = \Delta \tilde{r} / \gamma^{0.5}_{\rm meso}$,
the nonaffine displacement $\Delta \tilde{r}$ scaled by 
the local strain
$\gamma^{0.5}_{\rm meso}$.  The data shown are from one
experimental run, using portions of the data corresponding to
strain increments $\gamma_{\rm meso}$ = 0.01, 
0.02, 0.10, and 0.43.  These correspond
to time intervals $\Delta t=1.5$, 4.5, 16.5, and 45~s.
$P(\Delta s)$ is normalized so that
the integral over all vectors $\Delta \vec{s}$ is equal to 1,
similar to Ref.~\cite{Maloney08}.
For small strains, the curves have large exponential tails,
and for larger strain, the tails become smaller and appear more Gaussian.
The dashed line is Gaussian fit for $\gamma_{\rm meso}=0.43$.
For this sample,
the data and parameters are the same as Fig.~\ref{Trajectory}.
}
\end{figure}

To better understand the mean square displacement curves
(Fig.~\ref{MSD}), we wish to examine the data being averaged
to make these curves.  We do so by plotting the distributions
of displacements $\Delta \tilde{r}$ in Fig.~\ref{Pds}.  To better
compare the shapes of these distributions, we normalize these
displacements by the strain and thus plot $P(\Delta s)$ where
$\Delta s \equiv \Delta \tilde{r} / (\gamma^{0.5}_{\rm meso})$; this
normalization is motivated by the observation that at large
$\gamma_{\rm meso}$, we have $\langle \Delta \tilde{r}^2 \rangle \sim
\gamma_{\rm meso}$ (Fig.~\ref{MSD}) \cite{Maloney08}.
Furthermore, we normalize $P(\Delta s)$ so that the integral over
all vectors $\Delta \vec{s}$ is
equal to 1, similar to Ref.~\cite{Maloney08}.  Figure \ref{Pds}
shows that the distributions corresponding to small strains are
much broader than those corresponding to large strains.  For the
smallest strain, the distribution has a large exponential tail
over 3 orders of magnitude.  For larger strains ($\gamma_{\rm meso}
>0.1$), the curves are no longer exponential and the tails are
shorter, indicating fewer extreme events.  These curves appear
more like Gaussian distributions.  At the larger strain values
($\gamma_{\rm meso} > 0.1$),
the distributions collapse; this is the same strain regime for
which the mean square displacement becomes linear with
$\gamma_{\rm meso}$ [Fig.~\ref{MSD}(b)].
As the Ref.~\cite{Maloney08}
suggests, the exponential tail for small strains is similar to
what has been seen for individual plastic events
\cite{Tanguy06,Lemaitre07}, while the distributions for larger strains are
consistent with successive temporally uncorrelated plastic
events drawn from the exponential distribution. However,
it is possible that these events are spatially correlated,
which will be seen below.

\begin{figure}[htbp]
\includegraphics[scale=0.35, angle=-90 ]{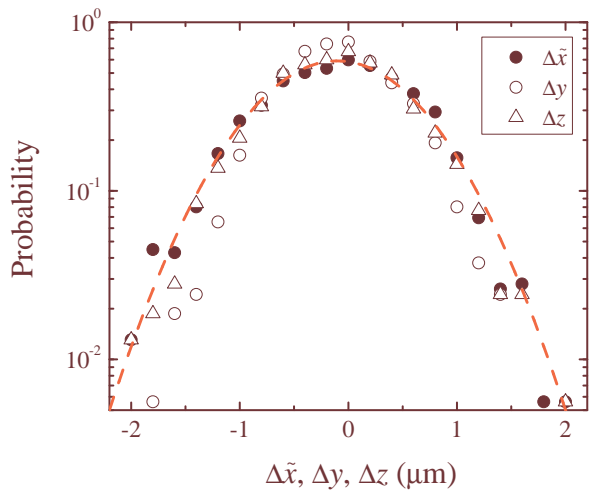}
\caption{\label{PDF_nonaffine}
(Color online)
Probability distribution functions for non-affine motions
in each direction,
$\Delta \tilde{x}$ (filled cycle),
$\Delta y$ (hollow circle)
and $\Delta z$ (hollow triangle).
The red curve is a Gaussian fit to the
$\Delta \tilde{x}$ data.
For this sample,
the parameters are the same as Fig.~\ref{Trajectory}
($\phi=0.51$,
$\gamma_{\rm meso}=0.43$,
$\bar{\dot{\gamma}}_{\rm meso}=0.0096$~s$^{-1}$,
$\Delta t = 45$~s).
}
\end{figure}

The mean square displacement data we have shown (Fig.~\ref{MSD})
treats all three directions equivalently, with the exception
that the $x$ displacements have had their nonaffine motions
removed.  However, the three directions are not equivalent
physically:  the $x$ direction corresponds to the shear-induced
velocity, the $y$ direction is the vorticity direction, and $z$
is the velocity gradient direction.  To look for differences in
motion between these three directions, we plot the probability
distribution of the displacements $[\Delta \tilde{x}, \Delta
y , \Delta z ]$, in Fig.~\ref{PDF_nonaffine}.  The three
distributions agree with each other, and in fact are symmetric
around the origin.  This suggests that the shear-induced
motions are isotropic.  Furthermore, they are well-fit by a
Gaussian, suggesting that the shear-induced motion liquefies
the sample (at least at the large $\gamma_{\rm meso}$ considered
for Fig.~\ref{PDF_nonaffine}).  This seems natural in the
context of jamming, where adding more strain moves the sample
farther from the jammed state \cite{Liu98}.
Of course, in our raw data the $\Delta x$ data show a
significant bias in the direction of the shear-induced velocity;
but it is striking that the non-affine displacements $\Delta
\tilde{x}$ show no difference from the displacements in $y$
and $z$, as also found in sheared colloidal glass
\cite{Besseling07,Yamamoto04}.

\begin{figure}[htbp]
\includegraphics[scale=0.40,angle=-90]{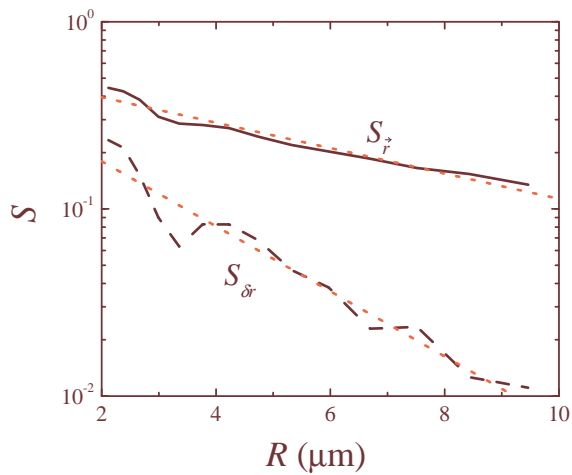}
\caption{\label{Correlation_S}
(Color online)
Spatial correlation functions $S_{\vec r}$ and $S_{\delta r}$
characterizing particle motion.  The data are the same as
Fig.~\ref{Trajectory} ($\phi=0.51$, 
$\gamma_{\rm meso}=0.43$ over 45~s of data, 
and $\bar{\dot{\gamma}}_{\rm
meso} = \gamma_{\rm meso}/\Delta t = 0.0096$~s$^{-1}$).
The solid and dashed lines correspond to $S_{\vec r}$ and $S_{\delta r}$
averaged over all the particles in the sample.
The dotted lines (red) correspond to exponential fits with length
scales $\xi_{\vec r} = 6.4$~$\mu$m and $\xi_{\delta r} =
2.5$~$\mu$m.
}
\end{figure}

Thus far, we have established that shear-induced particle
displacements are closely tied to the total applied strain
$\gamma_{\rm meso}$ [Fig.~\ref{MSD}(b)].  We then
introduced the nonaffine motion $\Delta \tilde{r}$ which we find
to be isotropic, on the particle scale:  individual particles are
equally likely to have shear-induced displacements in any
direction (Fig.~\ref{PDF_nonaffine}).
While the distributions of displacements are isotropic, this does
not imply that displacements are uncorrelated spatially.  To
check for this, we calculate two displacement correlation
functions as defined in
Ref.~\cite{Weeks07,Crocker00,Doliwa00} $:$
\begin{eqnarray}
S_{\vec{r}}(R, \Delta t) &=&
\frac{\langle \Delta \vec{\tilde{r}}_i \cdot \Delta \vec{\tilde{r}}_j
\rangle}{\langle
\Delta \tilde{r}^2 \rangle},\\
S_{\delta r}(R, \Delta t) &=&
\frac{\langle \delta \tilde{r}_i \delta \tilde{r}_j \rangle}{\langle
(\delta \tilde{r})^2 \rangle},
\end{eqnarray}
where the angle brackets indicate an average over all particles;
see Refs.~\cite{Weeks07,Doliwa00} for more details.  The mobility
is defined as $\delta \tilde{r} = |\Delta \tilde{r}| - \langle
|\Delta \tilde{r}| \rangle$, in other words, the deviation of the
magnitude of the displacement from the average magnitude of all
particle displacements.  The correlation functions are computed for
the nonaffine displacements, using $\Delta t = 45$~s to maximize
the ``signal''
(nonaffine displacements) compared to the ``noise'' (Brownian
motion within cages, on a shorter time scale).

The two correlation functions are shown in Fig.~\ref{Correlation_S}
for a representative data set. These functions are large at
short separations $R$ and decay for larger $R$, suggesting
that neighboring particles are correlated in their motion.
In particular, the vector-based correlation $S_{\vec r}$ has a
large magnitude at small $R$, showing neighboring particles have
strongly correlated directions of motion, even given that we are
only considering the nonaffine displacements. The two correlation
functions decay somewhat exponentially, as indicated by the
straight line fits shown in Fig.~\ref{Correlation_S}, with decay
constants $\xi_{\vec r} = 6.4$~$\mu$m$= 6.1a$ and $\xi_{\delta
r} = 2.5$~$\mu$m$=2.4a$ (in terms of the particle radius $a$).
The larger the slope, the more localized the correlation is.
$\xi_{\vec r}$ is similar to that found previously for supercooled
colloidal liquids, and $\xi_{\delta r}$ is slightly shorter than
the prior results \cite{Weeks07}.  Overall, these results confirm
that the shear-induced particle motion is spatially heterogeneous,
quite similar to what has been seen in unsheared dense liquids
\cite{Ediger96,Donati98,Weeks00,Weeks07,Kegel00,Ediger00,Doliwa00}
and granular materials \cite{Mehta08,Goldman06}.  The length scale
may be equivalent to the correlation length scale for fluidity
discussed in Refs.~\cite{goyon08,bocquet09}.  For example, an
experimental study of sheared polydisperse emulsions found a
fluidity length scale comparable to 1-2 droplet diameters near
the glass transition \cite{goyon08}.

Considering all of our data sets, we do not find a strong
dependence on either the strain rate $\dot\gamma$ or the total
strain $\gamma$ for the ranges we consider 
($\bar{\dot{\gamma}}_{\rm meso}
= 0.006 - 0.02$~s$^{-1}$, $\gamma_{\rm meso} = 0.3 - 1.6$).
We do not have a large amount of data with which to calculate the
correlation functions; unlike prior work, we cannot do a time
average \cite{Weeks07}.  If we use an exponential function to fit
our different data (different strains, strain rates, and volume
fractions),
the mobility correlation $S_{\delta r}$ yields a length
scale $\xi_{\delta r} \approx 1.8-3.9$~$\mu$m and the vector
correlation $S_{\vec{r}}$ yields a length scale $\xi_{\vec{r}}
\approx 3.2-7.5$~$\mu$m.  To check this, we also calculate
\begin{equation}
\xi' = \frac{ \langle R \cdot S(R) \rangle}{\langle S(R) \rangle}
\end{equation}
where the angle brackets indicate averages over $R$; for a perfect
exponential, we would have $\xi' = \xi$.  Using this method, we
find more consistent length scales of $\xi'_{\delta r} \approx
(3.0-4.0)a$ and $\xi'_{\vec{r}} \approx (3.5-4.3)a$.  Our data do
not suggest any dependence of these length scales on the control
parameters over the range we investigate.  Of course, as $\dot\gamma
\rightarrow 0$, we would expect to recover the original unstrained
sample behavior \cite{Yamamoto97}. Similar samples in this volume
fraction range were previously found to have length scales with
similar values, $\xi_{\delta r} \approx 4a-8a$ and $\xi_{\vec{r}}
\approx 6a$ \cite{Weeks07}. However, the time scales for this
motion are much longer than that for our sheared samples.

\subsection{Defining local plastic deformation}


We wish to look for spatially heterogeneous dynamics, that is,
how the shear-induced motion takes place locally and how particles
cooperate in their motion.  Several prior groups have examined local
rearrangements in simulations and experiments \cite{Falk98, Utter08,
Schall07,goyon08}, but have used differing ways to quantify the motion.
We will discuss those quantities together, and compare them using
our data.  We have two goals:  first, to understand how different
measures of local rearrangements reveal different aspects of the
motion; and second, to see if the spatial structure of rearranging
groups of particles exhibits any particular orientation with
respect to the shear direction.

For all of these definitions of rearranging groups of particles, it
is useful to define a particle's nearest neighbors.  Our definition
of a particle's nearest neighbors are those particles within
the cutoff distance $r_0$, set by the first minimum of the pair
correlation function $g(r)$.

\begin{figure}[htbp]
\includegraphics[scale=0.35, angle=-90 ]{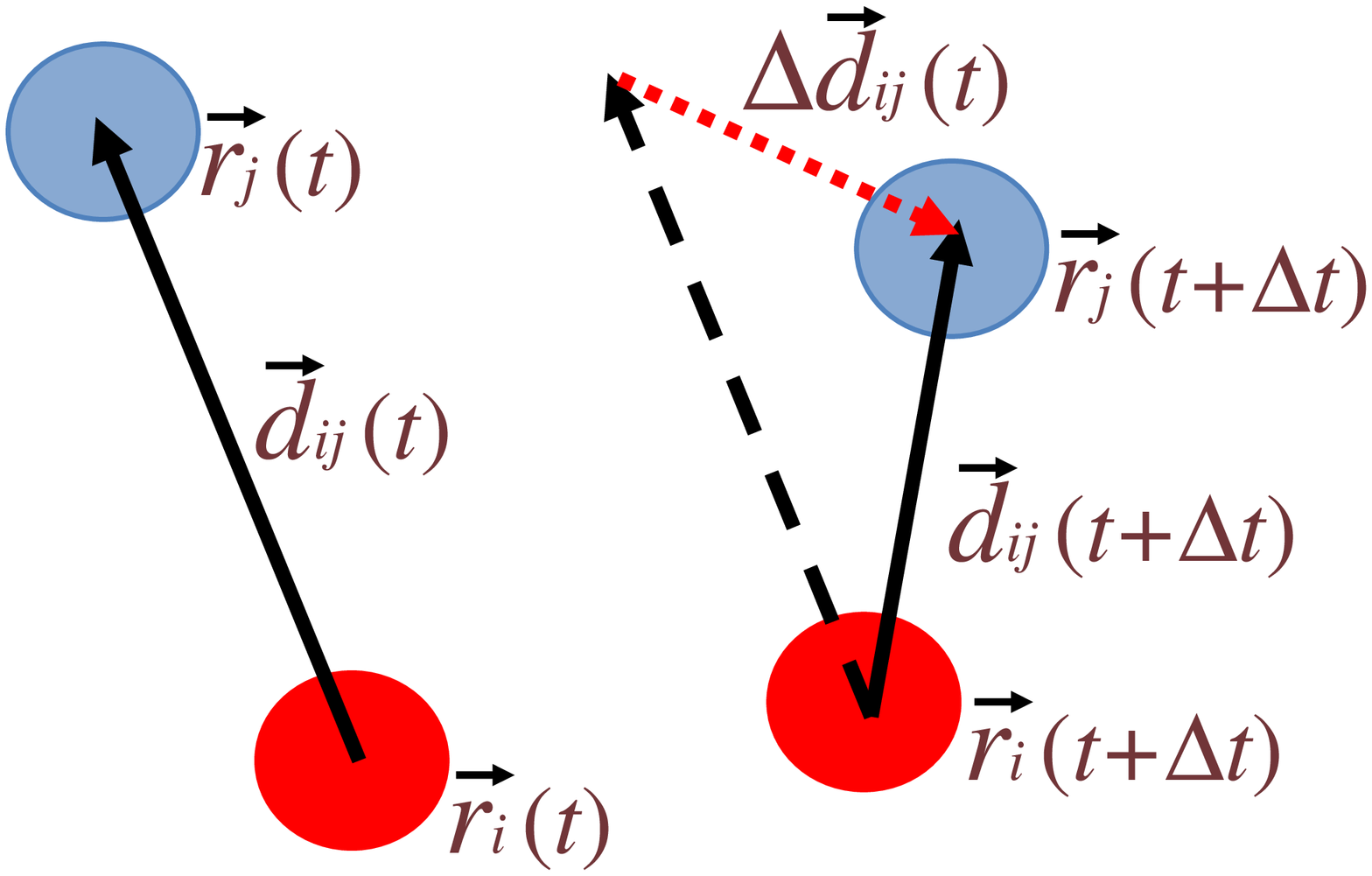}
\caption{\label{NeighborDemo}
(Color online)
Sketch illustrating the definition of the motion of the relative
position vector $\Delta \vec{d}_{ij}$ of two neighboring
particles, red and blue (dark and light gray).  $\vec{r}_{i}(t) $ represents
the position of particle $i$ at time $t$, $\vec{d}_{ij}(t)$
is the relative position between particles $i$ and $j$ at time
$t$, and $\Delta \vec{d}_{ij}(t)$ is how the vector $\vec{d}_{ij}$
changes over the time interval $(t,t+\Delta t)$.
}
\end{figure}

We start by quantifying the local strain seen by an individual
particle, which is based on the average motion of its neighbors.
For a particle $i$ with center at $\vec{r}_i(t)$, the relative
positions of its neighbors $j$ are $\vec{d}_{ij} (t) = \vec{r}_j
(t) - \vec{r}_i  (t) $.  These neighboring particles move, and
their motions over the next interval $\Delta t$ are given by
$\Delta \vec{d}_{ij}(t)=\vec{d}_{ij} (t+ \Delta t)-\vec{d}_{ij}
(t)$, as shown in Fig.~\ref{NeighborDemo}.

The strain tensor $\mathbb{E}_i$
for this region around particle $i$ is then determined
by minimizing the mean squared difference between
the actual relative motions $\Delta \vec{d}_{ij} (t) $
and that predicted by $\mathbb{E}_i$, in other words, choosing
$\mathbb{E}_i$ to minimize
\begin{equation}
   \label{dmin}
   D^2_{i,{\rm min}} =
   {\rm min} \Bigg[ {\Sigma_j} \ \bigg[ \Delta \vec{d}_{ij} (t)-
   \mathbb{E}_i \vec{d}_{ij} (t) \bigg ]^2 \ \Bigg].
\end{equation}
The error, $D^2_{i,{\rm min}}$, quantifies the plastic
deformation of the neighborhood around particle $i$,
after removing the averaged linear response
$\mathbb{E}_i \vec{d}_{ij} (t)$ \cite{Falk98}. 
Thus, $D^2_{i,{\rm min}}$ is
one way to quantify the local nonaffine rearrangement, ``local''
in the sense of an individual particle $i$ and its neighbors.
We term $D^2_{i,{\rm min}}$ the ``plastic deformation.''
Note that the sum is computed over the ten nearest particles
$j$ to particle $i$, otherwise the value of $D^2_{i,{\rm
min}}$ would depend on the number of neighbors.  In practice,
most of these neighboring particles are within 3.0~$\mu$m of
particle $i$, which is comparable to the first minimum of the
pair correlation function $g(r)$, which motivates our choice
of ten neighbors.

Of course, quite often $\mathbb{E}_i$ is different
from the overall strain over the imaged volume, which in
turn is different from the macroscopically applied strain.
In practice, given that the shear is applied in $x$ direction
with the velocity gradient along $z$, we only treat the $xz$
components of Eqn.~\ref{dmin}; that is, $\vec{d}_{ij}$ and
$\mathbb{E}_{ij}$ can be written as $:$
\begin{equation}
\label{ematrix}
\vec {d}_{ij} = \begin{bmatrix} x_{ij} \\
                                 z_{ij}
                 \end{bmatrix} ,
\mathbb{E}_i = \begin{bmatrix} \epsilon _i^{xx}  &  \epsilon _i^{xz} \\
                             \epsilon _i^{zx}  &  \epsilon _i^{zz}
             \end{bmatrix} .
\end{equation}


To better understand this local strain tensor $\mathbb{E}_i$,
we follow the method of Ref.~\cite{Schall07, Falk98}.  If the
particle-scale local strain was identical to the imposed
strain, we would expect $\epsilon_i^{xz}=\gamma_{\rm meso}$
and the other matrix elements to be zero.  We find that
these expectations are true on average (for example $\langle
\epsilon_i^{xz} \rangle = \gamma_{\rm meso}$) but that for
individual particles their local environment can be quite
different.  For each experiment, the distributions of all four
matrix elements have similar standard deviations, and examining
different experiments the standard deviations are between 24\% -
39\% of $\gamma_{\rm meso}$.

%
%
%
%
%
%
%
%
%
%
%

To quantify the measured particle-scale strain, we define the
``local strain''
\begin{equation}
\label{microstrain}
\gamma_{i,{\rm micro}} = \epsilon^{xz}_i
+ \epsilon^{zx}_i
\end{equation}
(using the definition of the strain
tensor which is related to $\mathbb{E}$ \cite{Landau}).
That is, this quantity is a local approximation to the strain
$(\frac{\partial u_x}{\partial z} + \frac{\partial u_z}{\partial
x} )$.  The local strain $\gamma_{i,{\rm micro}}$ is now a
second way to quantify the local rearrangement of the
neighborhood around particle $i$, in addition
to $D^2_{i,{\rm min}}$.  The diagonal elements, $\epsilon^{xx}_i$
and $\epsilon^{zz}_i$, relate to the dilation of the local
environment.  In particular, the local environment stretches by
a factor of $(1+\epsilon^{xx}_i)$ in the $x$ direction and likewise
$(1+\epsilon^{zz}_i)$ in the $z$ direction.  If these matrix
elements are zero, then the local environment remains the same;
positive matrix elements correspond to expansion and negative
matrix elements correspond to contraction.  We define the overall
dilation as $\delta e_i = (1+\epsilon^{xx}_i)(1+\epsilon^{zz}_i)-1$,
which is a third way to quantify the local rearrangement around
particle $i$.

A fourth way to consider local particle motion is the previously
defined nonaffine displacement, $\Delta \tilde{r}^2$.
The key difference is that $D^2_{\rm min}$,
$\gamma_{\rm micro}$, and $\delta e$ all are derived from the actual
particle motion $\Delta r$, whereas $\Delta \tilde{r}^2$
removes the motion caused by $\gamma_{\rm meso}$
(through Eqn.~\ref{nonaffineeqn}).

\begin{figure}[htbp]
\includegraphics[scale=0.38]{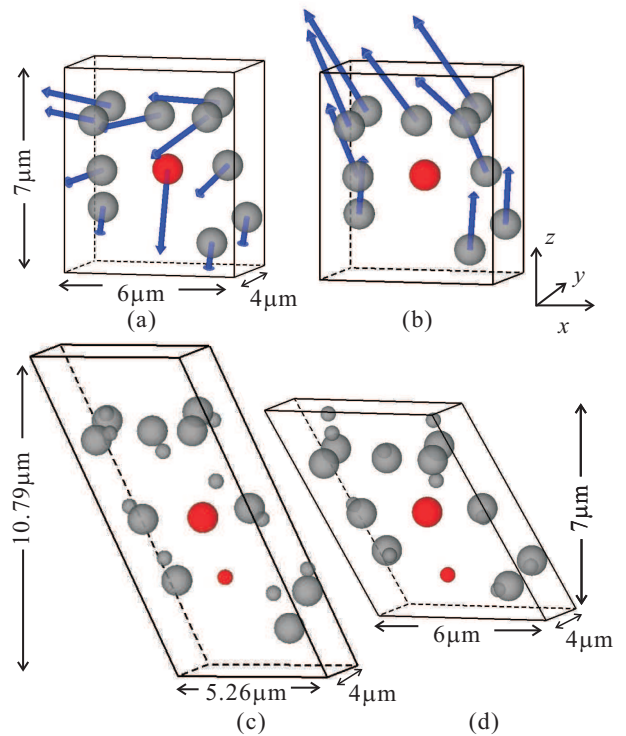}
\caption{\label{Demo_Deform}
(Color online.)
These four sketches show different portrayals of a particle with
unusual motion.  (a) Particle motion as seen in the laboratory
reference frame.  The arrows indicate displacement vectors.
(b) Similar to (a) except the motion are in the reference
frame where the central red (darker) particle is motionless.
(c) The large spheres correspond to the expected positions of
the particles, and the smaller spheres correspond to the actual
positions of the particles.  The distortion of the box and the
expected positions of the particles are calculated based on the
measured local strain tensor $\mathbb{E}_i$, where the
red (darker) particle is particle $i$.  For this local neighborhood, we
have $\gamma_{\rm micro} = 0.73$ and a dilation primarily in
the $z$ direction ($\epsilon^{xx}=-0.1, \epsilon^{zz}=0.5$, and
$\delta e=0.35$).
(d) The positions of the same particles after the strain has
been applied, where now the box represents the mesoscopic
strain $\gamma_{\rm meso} = 0.58$.  The large spheres
represent the expected positions if the motion was affine,
and the small spheres show the actual positions.  For all
panels, to show the displacements in three dimensions better,
the radii of the large spheres are 0.5 of the real scale.
The data correspond to a sample with $\phi=0.51$, $\gamma_{\rm
meso}=0.58$, $\bar{\dot{\gamma}}_{\rm meso}=0.014$~s$^{-1}$,
and $\Delta t = 40$~s.
}
\end{figure}

To demonstrate how neighboring particles rearrange and result in
larger values of these various parameters, Fig.~\ref{Demo_Deform}
shows an example using real trajectories.  The original positions of
the particles are shown, along with displacement vectors indicating
where they move after the sample is strained with $\gamma_{\rm
meso}=0.58$.  The overall strain is seen in that particles near the
top move farther than those at the bottom; however, the red (dark)
particle in the middle has an unusual motion, moving downward in
the $-z$ direction.  Figure \ref{Demo_Deform}(b) shows the motion
of the surrounding particles, as seen in the reference frame
attached to the red (dark) particle.  It is these
displacements that are used in the calculation Eqn.~\ref{ematrix}.
Figure \ref{Demo_Deform}(c) shows the predicted final
positions of the particles (drawn large) based on $\mathbb{E}_i$,
as compared to the actual final positions (drawn small); the red
(dark) particle is considered ``particle $i$''.  This local region
experiences both shear and a strong dilation in the $z$ direction,
both captured by $\mathbb{E}_i$.  The differences between the
predicted and actual final positions result in a moderately
large value of $D^2_{\rm min}=56$~$\mu$m$^2$.  In particular,
note that $D^2_{\rm min}$ is defined based on vectors pointing
from the red (dark) reference particle to the other particles,
and because the red particle moves downward, the vectors are all
greatly stretched and this increases $D^2_{\rm min}$.  Finally,
Fig.~\ref{Demo_Deform}(d) shows the positions of the same particles
after the strain has been applied, where now the box represents the
mesoscopic strain $\gamma_{\rm meso} = 0.58$.  The large spheres
represent the expected positions if the motion was affine, and
the small spheres show the actual positions.  Differences between
the expected and actual positions result in large values of the
nonaffine displacement $\Delta \tilde{r}$.  For the red (dark)
particle, $\Delta \tilde{r} = 2.97$~$\mu$m.  Overall, the anomalous
motion of the central particle $i$ is because under the large local
strain, this particle makes a large jump out of its original cage.
It is these sorts of unusual motions that result in large plastic
deformations within the sample.

%


\subsection{Collective particle motions}

\begin{figure}[htbp]
\includegraphics[scale=0.55]{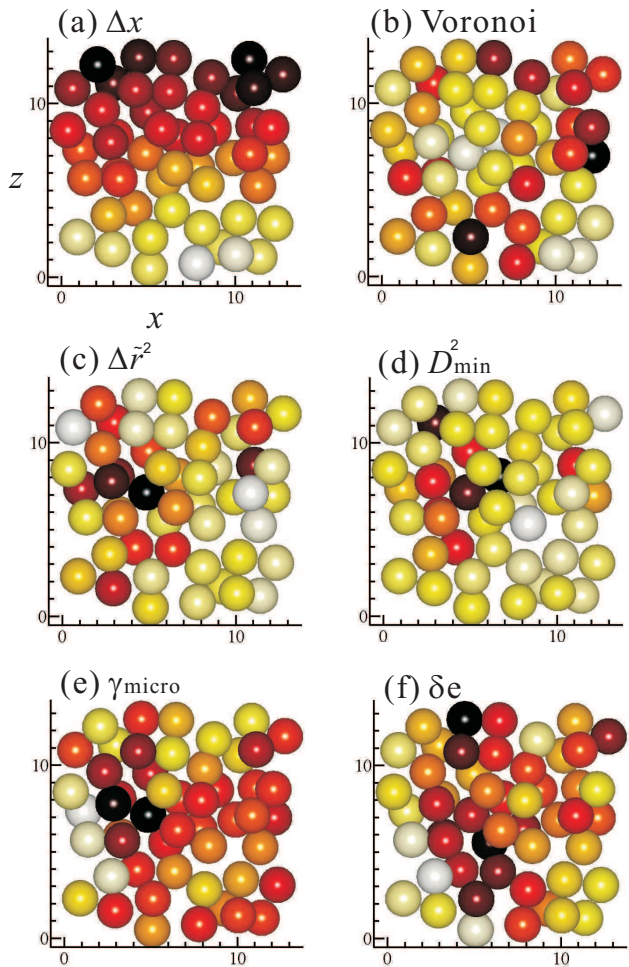}
\caption{\label{3Dyslice_Quantities}
A 3~$\mu$m thin cut through the
sample, showing the $xz$ shear plane; the axes are
as labeled in (a).  The particles are drawn in their location at
$t=0$, the start of this strain half-cycle.
(a) Darker particles have larger values of $\Delta x$, thus
indicating the applied strain.  The top particles (large $z$) are
moving left (dark colors) and the bottom particles are moving
right (light colors).
(b) Darker particles have larger
Voronoi volumes at $t=0$.
(c) Darker particles have larger values of
the non-affine motion $\Delta \tilde{r}^2$.
(d) Darker particles have larger values of
the plastic deformation $D^2_{\rm min}$.
(e) Darker particles have larger
values of the local strain $\gamma_{\rm micro}$.
(f) Darker particles have larger
values of the dilation $\delta e$.
The sample is the same as the data shown in
Fig.~\ref{Trajectory}; see that figure caption for details.
}
\end{figure}

To investigate the relationships between these quantities, a
3~$\mu$m-thin $y$ slice of a sample with volume fraction $\phi=0.51$
is shown in several ways in Fig.~\ref{3Dyslice_Quantities}.
In panel (a), the $x$ displacement is shown, making the strain
apparent.  Panel (b) shows the original Voronoi volumes for each
particle at $t=0$.  The Voronoi cell for each particle is defined as
the volume closer to the center of that particle than to any other
particle.  In subsequent panels, the darker colors indicate larger
local rearrangements, as measured by the non-affine displacement
$\Delta \tilde{r}^2$ [panel (c)], plastic deformation $D^2_{\rm
min}$ [panel (d)], local strain $\gamma_{\rm micro}$ [panel (e)],
and dilation $\delta e$ [panel (f)].  It can be seen that the
darker colored particles cluster together, indicating that for
each of these measures of local rearrangement, the motions are
spatially heterogeneous \cite{Schall07}.  This is a real-space
picture showing conceptually what is indicated by the correlation
functions in Fig.~\ref{Correlation_S}, that neighboring particles
have similar motions.  These pictures are qualitatively similar to
those seen for thermally-induced cage rearrangements in supercooled
liquids \cite{Kegel00,Weeks00,Donati98,Yamamoto98,Berthier04} and
glasses \cite{Courtland03,Lee02,Stevenson06,Goldman06,Kawasaki07}.
Furthermore, by comparing these images, it is apparent
that particles often have large values of several quantities
simultaneously; in particular compare panels (c) and (d), and panels
(e) and (f).  While the correspondence is not exact, it suggests
that all four of these ideas are capturing similar features
of rearranging regions.  However, it is also clear that there
are differences between (c) and (d) as compared to (e) and (f).
In the latter two panels, the region of high activity is spread
out over a larger area; more of the particles are deforming by
these measures at the same time.  Nonetheless, in all four cases,
the particles around $x \approx 2 - 4$~$\mu$m are experiencing
the most extreme deformations.

The Voronoi volume [Fig.~\ref{3Dyslice_Quantities}(b)] has
previously been found to be slightly correlated with particle motion
in unsheared colloidal supercooled liquids \cite{Weeks02,Conrad06};
that is, particles with large Voronoi volumes have more space to
move and thus are likely to have larger displacements than average.
Here it appears that there is no correlation between the Voronoi
volume and the particle motion, suggesting that for these strained
samples the local volume is not a crucial influence on the motion
\cite{Yamamoto04}.  This is probably because in the end, all
cages must rearrange for the strain to occur.

\begin{figure}[htbp]
\includegraphics[scale=0.50]{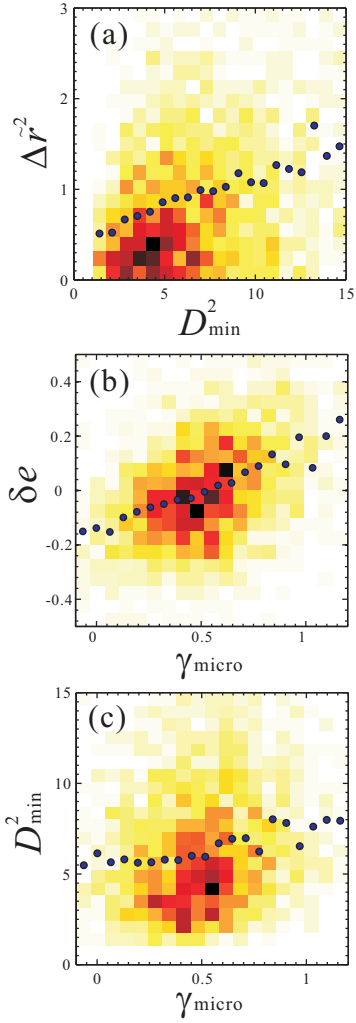}
\caption{\label{Quantities_Compare}
(Color online)
2D histograms of (a) $D^2_{\rm min}$ and $\Delta \tilde{r}^2$,
(b) $\gamma_{\rm micro}$ vs. dilation $\delta e$ and (c) $\gamma_{\rm
micro}$ and $D^2_{\rm min}$.  The sample has the same
parameters as Fig.~\ref{Trajectory}; see that caption for
details.  In these histograms, darker colors stand for the
larger probability.  The dotted lines are the average values
of the $y$ axis corresponding to each value on the $x$ axis.
The correlation coefficients between each pair of variables are
(a) $C=0.43\pm 0.11$, (b) $C=0.42\pm 0.09$, and (c) $C=0.17\pm
0.05$; see text for details.
}
\end{figure}

To demonstrate the relationships between the measures of plastic
deformation more quantitatively,
we compute 2D histograms comparing pairs of the variables,
shown in Fig.~\ref{Quantities_Compare}.  The darker
color indicates larger joint probability, and the dotted line
represents the mean value of the quantity on the vertical
axis corresponding to the quantity on the horizontal axis.
Figure~\ref{Quantities_Compare}(a) shows that on average,
particles with a large plastic deformation $D_{\rm min}^2$
are also much likelier to have a large nonaffine displacement
$\Delta \tilde{r}^2$.  This is suggested by the specific example
shown in Fig.~\ref{Demo_Deform}(c,d).
Similarly, Fig.~\ref{Quantities_Compare}(b) shows
that a particle's microscopic strain $\gamma_{\rm micro}$ is
well correlated with the dilation $\delta e$.  For these data,
the mesoscopic strain is $\gamma_{\rm meso}=0.43$; particles
with $\gamma_{\rm micro} < \gamma_{\rm meso}$ are more often
in local environments that contract ($\delta e < 0$), and
vice-versa.  As a contrast, Fig.~\ref{Quantities_Compare}(c)
shows a somewhat weaker correlation between $D^2_{\rm min}$ and
$\gamma_{\rm micro}$.  The Pearson correlation coefficients
between these quantities are $C(D^2_{\rm min},\Delta
\tilde{r}^2)=0.43 \pm 0.11$, $C(\gamma_{\rm micro},\delta
e)=0.42 \pm 0.09$, and $C(\gamma_{\rm micro},D^2_{\rm min})=0.17
\pm 0.05$.  The uncertainties are from the standard deviations
of the correlation coefficients from the nine different
experiments we conducted.

Overall, Fig.~\ref{Quantities_Compare} suggests that $D_{\rm
min}^2$ and $\Delta \tilde{r}^2$ both capture the idea of
plastic deformation \cite{Lemaitre07}.  The correspondence between these two
variables is nontrivial, given that $\Delta \tilde{r}^2$
is based on trajectories with the overall strain removed
(the strain computed from all observed particles), whereas
$D^2_{\rm min}$ only accounts for the very localized strain of
the neighboring particles.  In contrast, $\gamma_{\rm micro}$
and $\delta e$ are well-suited to examine particles moving
in atypical ways; typical particles have $\gamma_{\rm micro}
\approx \gamma_{\rm meso}$ and $\delta e=0$.  These two separate
ideas (plastic deformation and atypicality) are only weakly
correlated.  Other than the three specific comparisons shown
in Fig.~\ref{Quantities_Compare}, all other comparisons are
even less correlated ($|C| < 0.1$).  We also examined the
quantities $|\gamma_{\rm micro} - \gamma_{\rm meso}|$ and
$|\delta e|$ as ways to measure the deviations from typical
behavior; these quantities are also only weakly correlated to
the other measures of deformation.

\begin{figure}[htbp]
\includegraphics[scale=0.35, angle=-90]{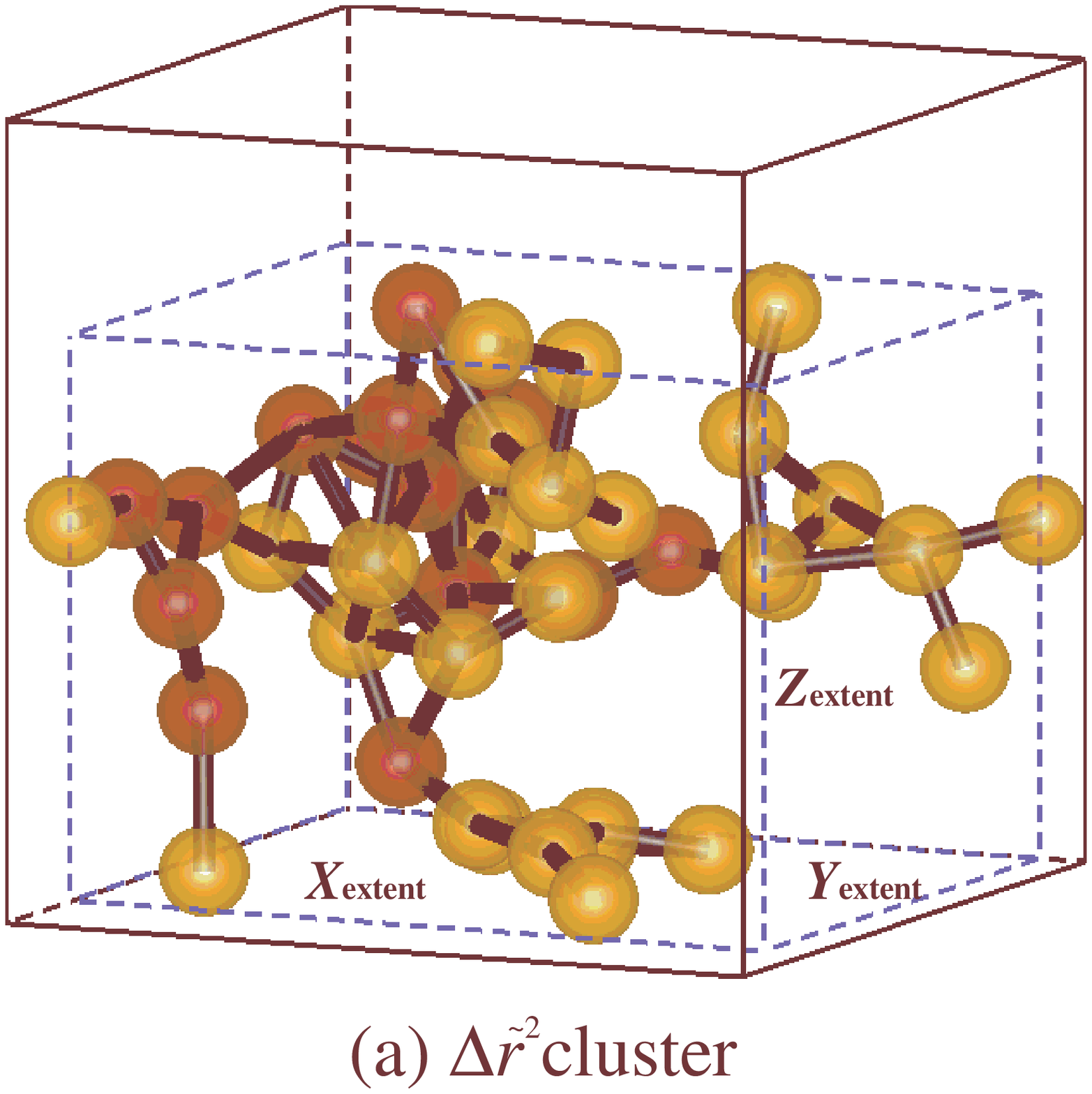}
\includegraphics[scale=0.35, angle=-90]{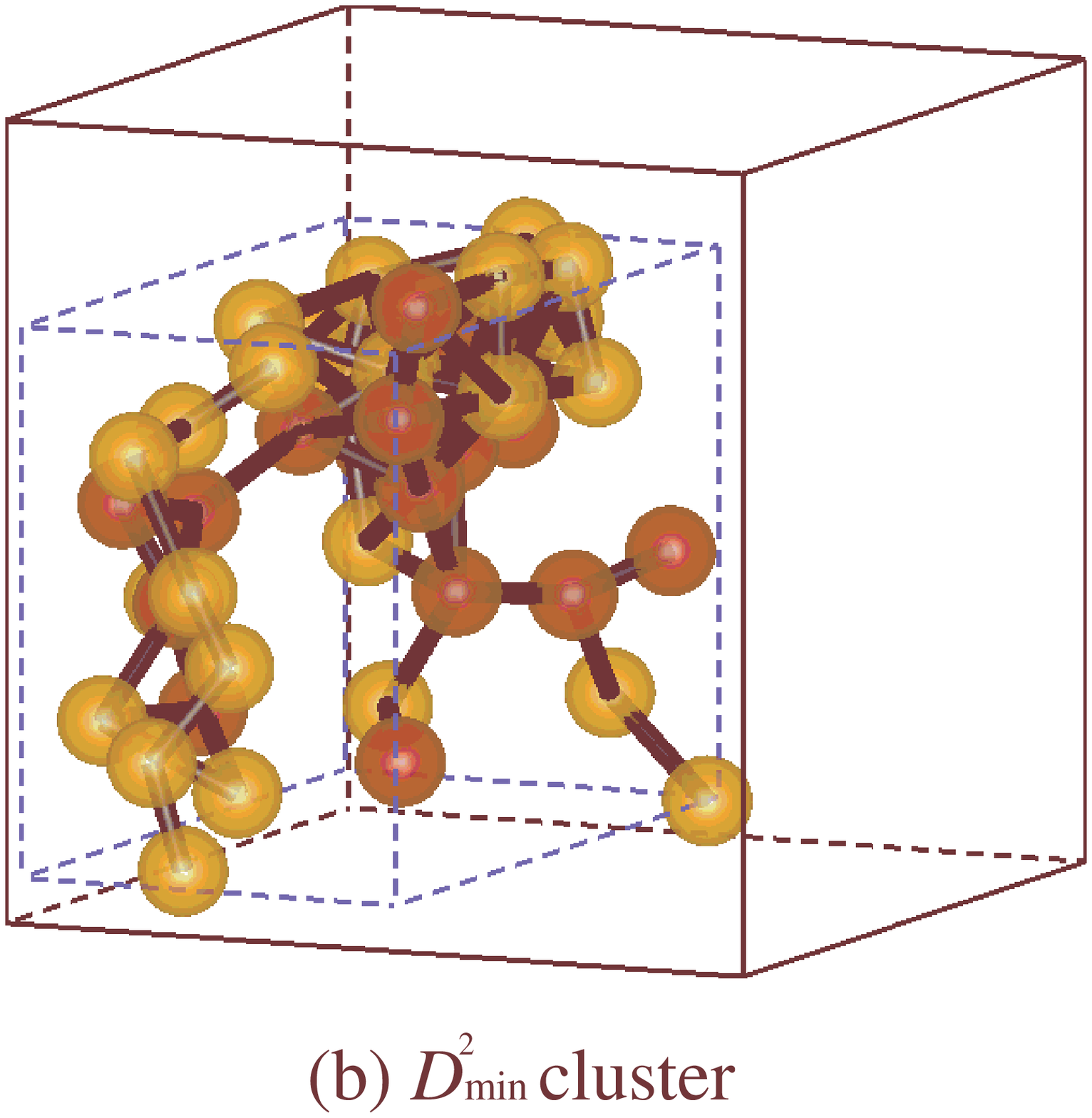}
\caption{\label{3Dcluster}
(Color online.)
A cluster of neighboring particles with large values
of (a) non-affine motion $\Delta \tilde{r}^2$ and (b)
plastic deformation $D^2_{\rm min}$.  To avoid biasing
the cluster shape, the particles are drawn from within
a 15$\times$15$\times$15~$\mu$m$^3$ cube (solid lines).
The dashed lines indicate the spatial extent of the two
cluster shapes in each direction.  The cluster in (a) is
$14 \times 12 \times 11$~$\mu$m$^3$ and the cluster in (b)
is $8 \times 14 \times 10$~$\mu$m$^3$.  The red particles are
common between the two clusters, and the black bonds indicate
neighboring particles; particle sizes are drawn 0.8 times the
actual scale, to make the connections more visible.
In each case, the particles shown are the
top 20\% for the parameter chosen.
The data corresponds to the same parameters as in
Fig.~\ref{Trajectory}; see that caption for details.
}
\end{figure}

We now return to the question of the isotropy of the motion.
Figure \ref{PDF_nonaffine} indicates that the distribution of
all particle motions is isotropic, but it is possible that the
spatially heterogeneous groups of highly mobile particles
shown in Fig.~\ref{3Dyslice_Quantities} are themselves
oriented along a preferred direction.  To investigate the 3D
structures of these relaxation regions, we quantify the sizes
of these active regions in the $x$, $y$ and $z$ directions.
To start, we define connected neighboring particles as those
with separations less than $r_0$, the distance where pair
correlation function $g(r)$ reaches its first minimum.
(Note that this is slightly different from the neighbor
definition used for Eqn.~\ref{dmin}; see the discussion
following that equation.)  For a given quantity, we consider
active particles as those in the top 20\% of that quantity,
similar to prior work \cite{Donati98,Weeks00,Courtland03,Yamamoto97}.
We then define the active region as a cluster of connected
active particles.  For example, Fig.~\ref{3Dcluster} shows
a cluster of particles with large non-affine displacements [panel
(a)] and a cluster with large plastic deformations [panel (b)].
Each cluster is drawn from the same data set, and the particles
drawn in red (darker) are common to both clusters.  (Note that
clusters drawn based on $\gamma_{\rm micro}$ and $\delta e$ are
smaller.  In Fig.~\ref{3Dyslice_Quantities}(e,f), more regions
have large values of these parameters, but the top 20\% most
active are not clustered to the extent they are in
Fig.~\ref{3Dyslice_Quantities}(c,d).)

We wish to understand if the shapes of such clusters show a
bias along any particular direction.  It is important that the
experimental observation volume not bias the result, so from within
the observed 3D volume we consider only particles that start within
an isotropic cube of dimensions 15$\times$15$\times$15~$\mu$m$^3$.
The size of this cube is chosen to be the largest cube for which
all the particles are within the optical field of view for the
full half-cycle of shear for all experiments.  (See the related
discussion at the end of Sec.~\ref{method}.)  Within this isotropic
volume, we consider the largest cluster from each experiment and
for each considered variable.  We then define the extent of that
cluster in each direction from the positions of each particle
within the cluster as $x_{\rm extent} = \max(x_i) - \min(x_i)$
and similarly for $y$ and $z$.


\begin{figure}[htbp]
\includegraphics[scale=0.55, angle=-90]{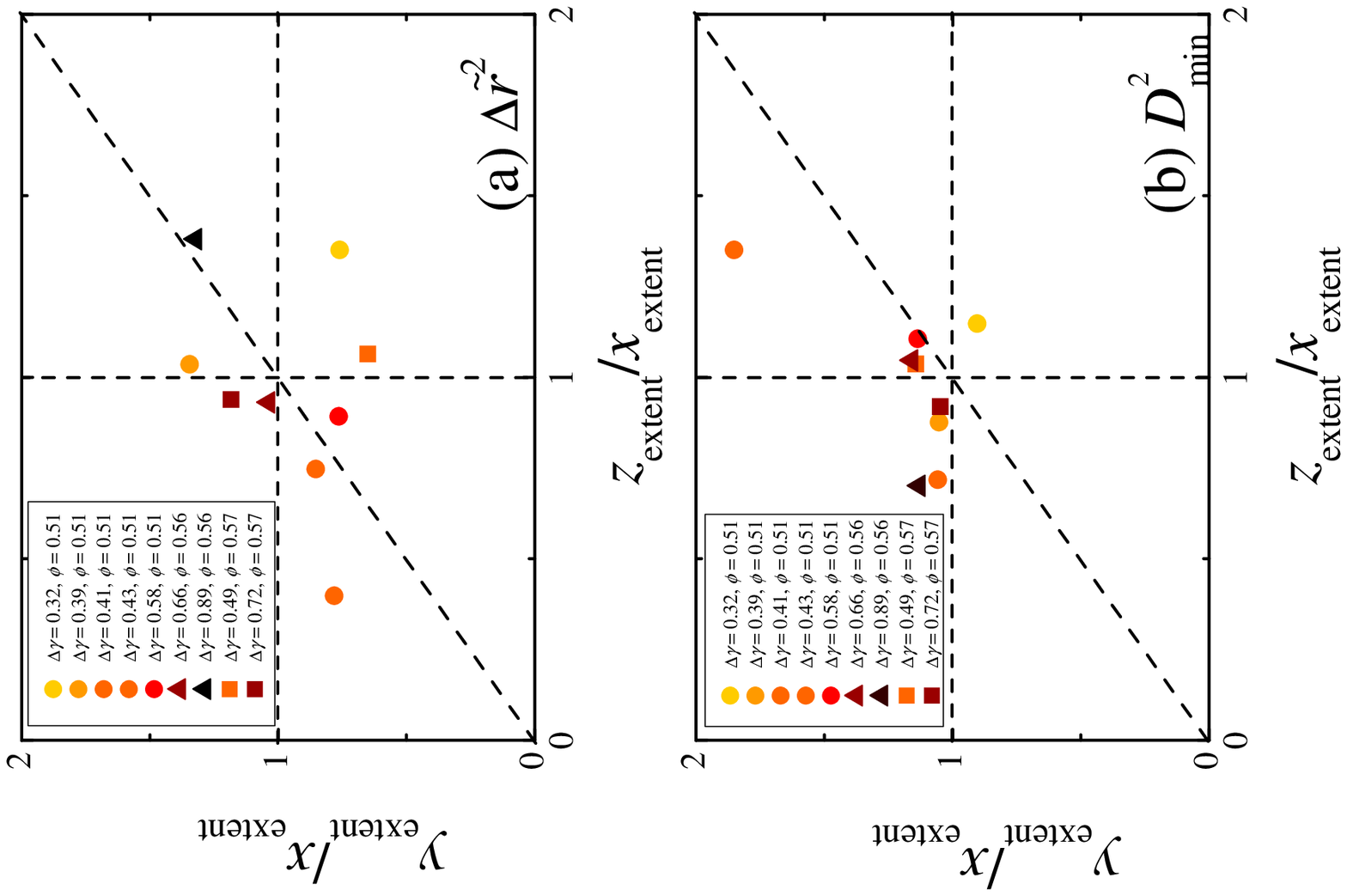}
\caption{\label{ClusterShape}
Comparison between the cluster
extent in $x$, $y$ and $z$, based on different variables.
(a) Particles with the largest non-affine motion $\Delta \tilde{r}$.
(b) Particles with the largest plastic deformation $D^2_{\rm min}$.
The symbols represent different volume fractions:
$\phi=0.51$ (circles), 0.56 (triangles) and 0.57 (squares).
Colors (or grays from light to dark) represent different accumulative strains
$\gamma_{\rm meso}=0.3 - 0.9$.
The clusters are comprised of the top
20\% of the particles with the given characteristic.
}
\end{figure}

Anisotropic cluster shapes would be manifested by systematic
differences in the relative magnitudes of $x_{\rm extent}$,
$y_{\rm extent}$, and $z_{\rm extent}$.  We compare these
in Fig.~\ref{ClusterShape} for clusters of particles with
large non-affine motion $\Delta \tilde{r}^2$ [panel (a)] and
large plastic deformation $D^2_{\rm min}$ [panel (b)].  The
comparison is made by using the ratios $y_{\rm extent}/x_{\rm
extent}$ and $z_{\rm extent}/x_{\rm extent}$, thus normalizing
the extent in the $y$ and $z$ directions by that of the shear
velocity direction $x$.  Thus, if a cluster has the same extent
in $x$ and $y$, $y_{\rm extent}/x_{\rm extent}$ should be equal
to $1$, along the vertical dashed line.  Similarly, for the same
extent in $x$ and $z$, points should be along the horizontal
dashed line with $z_{\rm extent}/x_{\rm extent}=1$.  If the
extent is the same for $y$ and $z$, the points should be along
the diagonal line with $y_{\rm extent}/x_{\rm extent}=z_{\rm
extent}/x_{\rm extent}$.  For an isotropic cluster with same
size in all dimensions, the point should be in the center (1,1).

As shown in Fig.~\ref{ClusterShape}, for all of our data, we
find no systematic anisotropy; the cluster extent ratios are
mostly clustered around the isotropic point (1,1). Due to
random fluctuations, no cluster is perfectly isotropic, yet
the points seem fairly evenly distributed around the three
dashed lines.  Thus, while the shear-induced rearrangements take
place in localized regions (Fig.~\ref{3Dyslice_Quantities}),
the data indicate that these regions on average have
no directional bias.  This seems true for
both the nonaffine displacements $\Delta \tilde{r}^2$
in Fig.~\ref{ClusterShape}(a) and the plastic deformation
$D^2_{\rm min}$ in Fig.~\ref{ClusterShape}(b).

\section{Discussion}
\label{Discussion}

We examined the microscopic plastic deformations occurring in
several sheared dense colloidal suspensions.  Our first main
observation is that on average, individual particles have no
bias in their direction of motion, other than that trivially
imposed by the strain.  When this imposed motion is removed from
the particle trajectories, the remaining shear-induced motion
is isotropic:  particles are equally likely to move in any
direction.  Our second main observation is on the shape of
groups of particles undergoing plastic rearrangements.  There are
several ways to determine which particles are rearranging, and
we have shown that all of these are useful for highlighting local
regions of deformation.  Furthermore, the shapes of these regions
are also isotropic.  However, we cannot rule out that with more data and
subtler analysis, we might find anisotropies in particle motion
\cite{Furukawa09}.

In our results, we find little dependence on the overall volume
fraction $\phi$, total strain, or strain rate.  For the volume
fraction, all of our samples are dense liquids with $\phi <
\phi_G$.  At significantly lower volume fractions, presumably
particles would not be caged and the shear-induced rearrangements
might be quite different \cite{Pine05}.  At higher volume
fractions $\phi > \phi_G$, prior work has seen similar results
\cite{Besseling07,Schall07} although not examined the shapes of
the rearranging regions in detail.  It is possible that results in
glassy samples might be different, given that near $\phi_G$ slight
changes in volume fraction have large consequences \cite{cheng02},
but we have not seen clear evidence of that in our data.  For the
total strain, we have not examined a wide range of parameters.
In all cases, we are studying sufficiently large enough strains to
induce irreversible, plastic rearrangements.For the strain rates,
all of our strain rates are fast enough such that the modified
Peclet number $Pe^{\star}$ is at least 7, so that thermally induced
diffusive motion is less relevant.  It is likely that at slower
strain rates (lower Peclet numbers), different behavior would be
seen \cite{Yamamoto97}.

Previous work \cite{Ackerson88, Ackerson90, Lequeux98, Haw98} found
that oscillatory shear can induce crystallization of concentrated
colloidal suspensions.  The `induction time' of this crystallization
is strain dependent:  a larger strain amplitude results in shorter
induction time.  In our experiments, we studied only a limited
number of oscillations, and our strain amplitude $\sim$1.  We did
not observe crystallization in any of our experiments.  It is likely
that were we to continue our observations for much longer times, we
could see the onset of shear-induced crystallization, and so we note
that our experiments are probably studying a non-equilibrium state.
Additionally, Fig.~\ref{Gammadot} shows that our strain rate takes
a while to stabilize after flow reversal, which again suggests
that our results are not in steady-state.  Thus, it is possible
that our primary observation, that the shear-induced particle
rearrangements are isotropic in character, is limited only to the
transient regime we observe.  It is still intriguing that in this
regime, particle motion is so isotropic.  For example,
Fig.~\ref{Gammadot} shows that the sample takes a while to
requilibrate after shear reversal, yet there is no obvious
signature of this in the particle motion or the configurations of
the particles.  Likewise, presumably the long-term
crystallization will be caused by anisotropic motion (and result
in further anisotropic motion), but no signs of this are present
in the early-time amorphous samples we study.  It would be
interesting to conduct longer-term experiments to relate the
particle rearrangements to those resulting in crystallization.
Alternatively, it would be also interesting to use a cone-and-plate
geometry shear cell capable of indefinitely large strains
\cite{Besseling09}, to reach the steady-state constant shear regime.


\section{Acknowledgments}
\label{Acknowledgements}

We thank R.~Besseling, J.~Clara Rahola, and V.~Prasad for
helpful discussions.  This work was supported by the National
 Science Foundation (DMR-0603055 and DMR-0804174).

\end{document}